\documentstyle[11pt,epsfig]{article}

\def\dis{\displaystyle}
\def\le{\left(}
\def\ri{\right)}

\def\no{\nonumber}

\def\f12{\frac{1}{2}}

\def\pd{\partial}

\def\L{\lambda}
\newcommand{\Li}{\mathop{\mathrm{Li}}\nolimits}

\begin{document}
\begin{titlepage}
\vskip 2cm
\begin{center}
{\Large \bf Transformations  of triangle ladder diagrams}\\
\vskip 1cm  
Igor Kondrashuk and Alvaro Vergara \\
\vskip 5mm  
{\it  Departamento de Ciencias B\'asicas, Universidad del B\'\i o-B\'\i o, \\
Campus Fernando May, Casilla 447, Chill\'an, Chile} \\
\end{center}
\vskip 2cm
\begin{abstract}
It is shown how dual space diagrammatic representation of momentum integrals corresponding to triangle ladder diagrams 
with an arbitrary number of rungs can be transformed to half-diamonds. In paper arXiv:0803.3420 [hep-th] the half-diamonds 
were related by conformal integral substitution to the diamonds which represent the dual space image of 
four-point ladder integrals in the four-dimensional momentum space. Acting in the way described in the present paper we do not need to 
use the known result for diamond (four-point) diagrams as an external input in deriving relations of arXiv:0803.3420 [hep-th], 
however, that result for the diamond diagram arises in the present proof as an intermediate consequence in a step-by-step diagrammatic transformation 
from the triangle ladder diagram to the half-diamond diagrams.
\vskip 1cm
\noindent Keywords: UD functions.
\end{abstract}
\end{titlepage}

\section{Introduction}

Triangle ladder diagrams  were initially investigated in the momentum space in Refs.\cite{Davydychev:1992xr,Usyukina:1992jd,Usyukina:1993ch}.
The result is  UD functions. In the momentum space integration is done over internal momenta that run in loops of the graph.  
However, the same diagram can be considered in the position space with integrals taken over coordinates of internal vertices in the graphs \cite{Kondrashuk:2008ec,Kondrashuk:2008xq} .
The results of the integrations in the position space and in the momentum space are related by Fourier transform. 
In Refs. \cite{Kondrashuk:2008ec,Kondrashuk:2008xq} the position space representation of the ladder diagrams has been investigated and it has been found that 
the result is the same UD functions. This means that the Fourier transform of the UD function are the 
same UD functions, and the correspondence is one-to-one, namely the UD function with number $n$ transforms to the UD function 
with number $n$ \cite{Kondrashuk:2008ec,Kondrashuk:2008xq}.  These relations follow ST identity  \cite{Slavnov:1972fg} - \cite{Zinn-Justin:1974mc} which was studied 
for  ${\cal N} = 4$ supersymmetric Yang-Mills theory in Refs. \cite{Cvetic:2004kx} - \cite{Kondrashuk:2000br}.

In deriving relations of Ref.\cite{Kondrashuk:2008xq} the dual graphical representation of the four-point ladder diagram in the form of ``diamond'' found in 
Ref.\cite{Isaev:2003tk,Drummond:2006rz}  has significantly been used as an external input
in order to relate half-diamonds to the UD functions  $\dis{\Phi^{(n)}(x,y)}.$  In this letter we show how to perform analysis of Ref.\cite{Kondrashuk:2008xq} 
starting with the three-point ladder diagrams and transforming them via dual space images to the half-diamonds. 
Passing several steps in the transformation procedure from the ladder diagrams to the half-diamond diagrams, we obtain the 
diamond diagrams at an intermediate step. However, we do not use the known result of Refs.\cite{Isaev:2003tk,Drummond:2006rz} for the diamond diagram. 
In contrary, that result is reproduced via dual space image of the triangle ladder diagram in the momentum representation.

The paper is organized as follows. In Section 2 the iterative definition of UD functions is recalled. In Section 3 dual space is reviewed and dual space representation of 
momentum triangle ladder diagrams is considered as a particular example. In Section 4 conformal substitution is analysed for the most simple example. In Section 5 the conformal 
substitution in the ladder integrals is applied to an arbitrary number of rungs. In Conclusion we write the formula for the Fourier 
transform of UD functions that follows the equations derived in Ref. \cite{Kondrashuk:2008xq} and discuss its particular features.

\section{Definition of UD functions}

In Refs. \cite{Usyukina:1992jd,Usyukina:1993ch} the definition of the UD functions has been done in the momentum space. 
\begin{figure}[htb]
\begin{minipage}[h]{1.0\linewidth}
\centering\epsfig{file=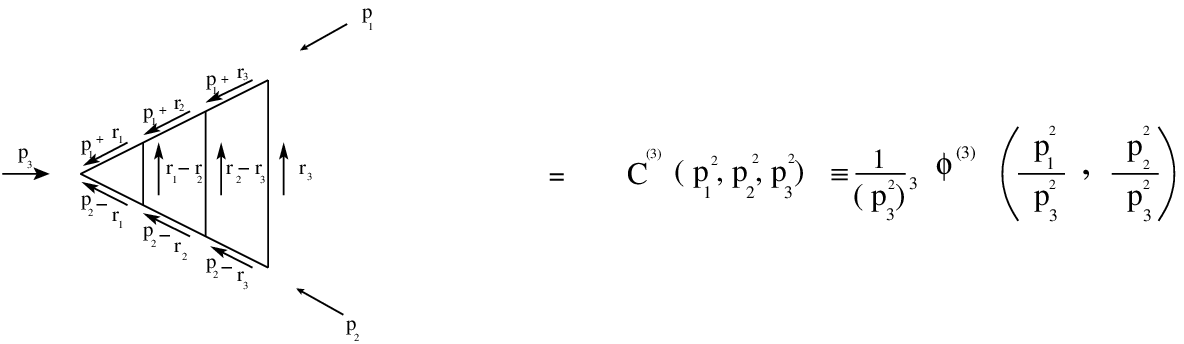,width=\linewidth}
\end{minipage}
\caption{\footnotesize Momenta in three-rung diagram}
\label{di1}
\end{figure}
For example, the result for the three-rung diagram presented in Fig.(\ref{di1}) is a function $C^{(3)}(p_1^2,p_2^2,p_3^2)$ 
of three independent Lorentz-invariant variables $p_1^2,p_2^2,p_3^2,$ constructed from the external momenta of the diagram 
which satisfy the conservation law $p_1 + p_2 + p_3 = 0.$  A useful parametrization for them is 
\begin{eqnarray}
p_1 = q_3 - q_2,\no\\ 
p_2 = q_1 - q_3,\no\\
p_3 = q_2 - q_1. \label{param}
\end{eqnarray}
Three four-dimensional vectors $q_1,~q_2,~q_3$ are independent \cite{Davydychev:1992xr,Usyukina:1992jd,Usyukina:1993ch}.

\begin{figure}[htb]
\begin{minipage}[h]{1.0\linewidth}
\centering\epsfig{file=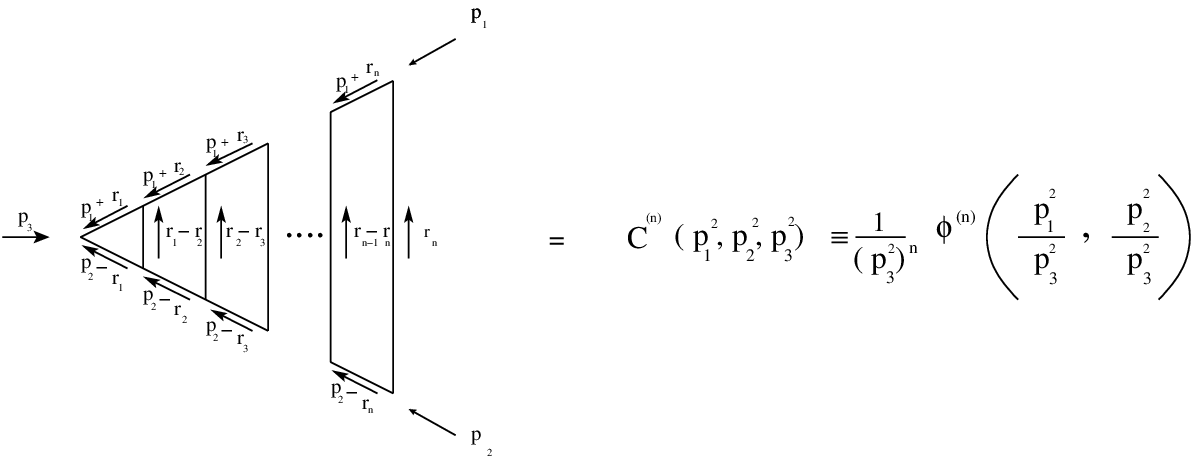,width=\linewidth}
\end{minipage}
\caption{\footnotesize Momenta distribution in $n$ rung diagram}
\label{di2}
\end{figure}

A ladder diagram with an arbitrary number of rungs is represented in Fig.(\ref{di2}), and the result for $n$ rung diagram 
$C^{(n)}(p_1^2,p_2^2,p_3^2)$ is expressed in terms of UD function with number $n.$ 
All the functions $\dis{\Phi^{(n)}\le \frac{p_1^2}{p_3^2}, \frac{p_2^2}{p_3^2} \ri}$ were calculated in 
Refs. \cite{Davydychev:1992xr,Usyukina:1992jd,Usyukina:1993ch} explicitly.

The iterative definition can be seen in Fig.(\ref{di2}) and is given by the equation 
\begin{eqnarray}
C^{(n)}(p_1^2,p_2^2,p_3^2) =  \int~d^4r_n~\frac{C^{(n-1)}((p_1 + r_n)^2,(p_2 - r_n)^2, p_3^2 )}{(p_1 + r_n)^2 (p_2 - r_n)^2 r_n^2}.\label{iter}
\end{eqnarray}

\section{Dual representation of the momentum integrals}

In Introduction the relation between the position space representation (p.s.r.) of some diagram and its momentum space representation (m.s.r.) is explained.
In addition to these representations  a dual space representation (d.s.r.) \cite{Kazakov:1986mu,Isaev:2003tk,Drummond:2006rz} exists. 
The dual space is not a physical space. However, it is useful mathematical construction that in case of ladder diagrams helps to find the relation between results of calculation in the p.s.r. 
and  in the m.s.r. for the same ladder graph. 
\begin{figure}[ht]
\begin{minipage}[h]{1.0\linewidth}
\centering\epsfig{file=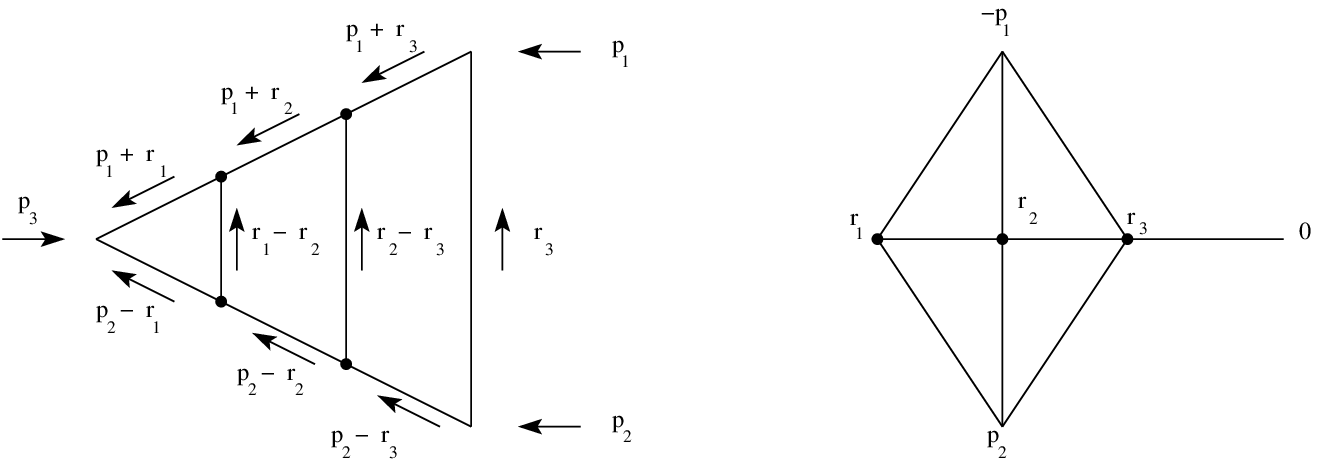,width=\linewidth}
\end{minipage}
\caption{\footnotesize The d.s.r. of the momentum space integral of the l.h.s. is on the r.h.s. of this picture}
\label{di3}
\end{figure}

In the d.s.r. of the integrals in the momentum space  momenta that run in loops of any Feynman diagram 
are treated as coordinates of internal vertices of a dual diagram in an auxiliary ``position'' space. 
The line that connects two vertices reproduces a propagator in the integrand of the momentum space integral. 
In the massless case the line which connects any two such vertices represents the inverse square of the interval 
between these two vertices. This precisely reproduces the massless propagators in the momentum space integrand (\ref{iter}). 
For example, the dual representation of the momentum integral for the three-rung ladder diagram of Fig.(\ref{di1}) 
is depicted in Fig.(\ref{di3}).  Graphically, the dual space diagram is another schematic representation of the same momentum 
integral. 
\begin{figure}[ht]
\begin{minipage}[h]{1.0\linewidth}
\centering\epsfig{file=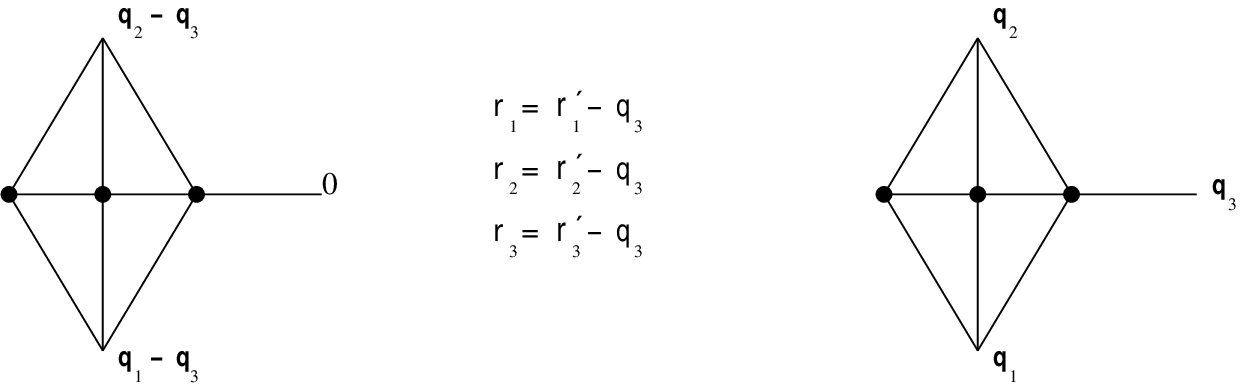,width=\linewidth}
\end{minipage}
\caption{\footnotesize Shift of variables of integration}
\label{di4}
\end{figure}

\begin{figure}[ht]
\begin{minipage}[h]{0.7\linewidth}
\centering\epsfig{file=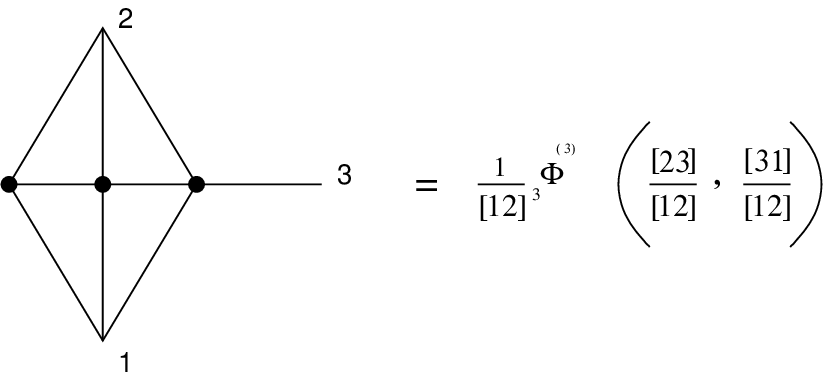,width=\linewidth}
\end{minipage}
\caption{\footnotesize Dual representation of three-rung diagram of Fig.(1)}
\label{di5}
\end{figure}

After re-parametrization according to Eqs.(\ref{param}) we make a shift of the integration variables $r_1,r_2,r_3$  
of the l.h.s. of Fig.(\ref{di3}) by the external value $q_3.$  As the result of this shift we obtain the r.h.s. of Fig.(\ref{di4}).
The new variables of integrations (after the shift) $r'_1,r'_2,r'_3$ are related to the initial variables
as  $r_1 = r'_1 - q_3,$ $r_2 = r'_2 - q_3,$ and $r_3 = r'_3 - q_3.$  The l.h.s. of Fig.(\ref{di4}) and  the r.h.s. of Fig.(\ref{di4})
are equal, and they are equal to the l.h.s. of Fig.(\ref{di5})  in which $N=1,2,3$ stands for $q_N=q_1,q_2,q_3.$
In this notation and in according to the definition of three-rung UD function of Fig.(\ref{di1}) the result for the diagram of 
Fig.(\ref{di5}) should be written as 
\begin{eqnarray*}
\frac{1}{[12]^3} \Phi^{(3)}\le \frac{[23]}{[12]},\frac{[31]}{[12]}\ri.
\end{eqnarray*}
In this paper concise notation $[Nq]= (r_N - q)^2$ of Ref.\cite{Cvetic:2006iu} for squares of intervals between two points of 
four-dimensional space is used.

However, the diagram of Fig.(\ref{di5}) in the dual space can be viewed as a diagram in the position space generated for another field theory.
The letters $q_1,q_2,q_3$ corresponding to the external points of the d.s.r. can be replaced with 
the letters $x_1,x_2,x_3$ corresponding to any three distinct points in the position space. The important feature 
of the construction is that in four space-time dimensions the propagator of scalar massless
field  between two points in the position space is square of the inverse interval between these two points. 
This happens in four dimensions only \cite{Kazakov:1984bw,Vasil}.
This correspondence allows the interpretation the dual space diagram as a position space diagram with massless scalar propagators.
The concise notation of  Ref.\cite{Cvetic:2006iu} for the space-time intervals is used in the rest of the paper, namely   
$[Nx]= (x_N - x)^2$  and $[12]= (x_1 - x_2)^2 ,$ that is, $N=1,2,3$ stands for $x_N=x_1,x_2,x_3.$

\section{Conformal substitution}

\begin{figure}[ht]
\begin{minipage}[h]{0.5\linewidth}
\centering\epsfig{file=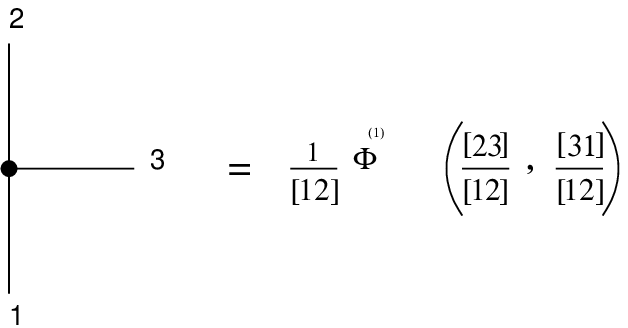,width=\linewidth}
\end{minipage}
\caption{\footnotesize First UD function in the position space}
\label{di6}
\end{figure}

Now we look at the diagram of the l.h.s. of  Fig.(\ref{di5}) as at a diagram in the position space. Our purpose is to transform this diagram 
to the half-diamond form of Ref.\cite{Kondrashuk:2008ec,Kondrashuk:2008xq}. An important tool to reach this purpose is a conformal substitution in the integrands. 
To demonstrate how to use it, we can consider the simplest example of the first UD function depicted in Fig.(\ref{di6}).
This picture corresponds to the integral definition of the first UD function \footnote{Our definition for UD functions is  $\Phi_{New}^{(L)} = (\pi^2)^L\Phi_{Old}^{(L)},$ 
where  $\Phi_{New}^{(L)}$ is  $\Phi^{(L)}$ of this paper, and   $\Phi_{Old}^{(L)}$ is the original UD function $\Phi^{(L)}$ 
of Refs. \cite{Usyukina:1992jd,Usyukina:1993ch}.} of Ref.\cite{Usyukina:1992jd}, 
\begin{eqnarray*}
J(1,1,1) = \int~d^4x~ \frac{1}{(x - x_1)^2 (x - x_2)^2 (x - x_3)^2} \equiv \\ 
\equiv  \int~d^4x~ \frac{1}{[1x][2x][3x]} = \frac{1}{[12]} \Phi^{(1)}\le \frac{[23]}{[12]},\frac{[31]}{[12]}\ri 
= \frac{1}{[31]} \Phi^{(1)}\le \frac{[12]}{[31]},\frac{[23]}{[31]}\ri.
\end{eqnarray*}
The conformal substitution for each vector of the integrand (including the external vectors) is 
\begin{eqnarray}
x_\mu = \frac{x'_\mu}{{x'}^2}, ~~~~ {x_1}_\mu = \frac{{x'_1}_\mu}{{x'_1}^2} \Rightarrow  [x1] = \frac{[x'1']}{[x'][1']};~~~~
[x] \equiv x^2,~~~~[1] \equiv x_1^2   \label{CT} 
\end{eqnarray}

\begin{figure}[ht]
\begin{minipage}[h]{0.7\linewidth}
\centering\epsfig{file=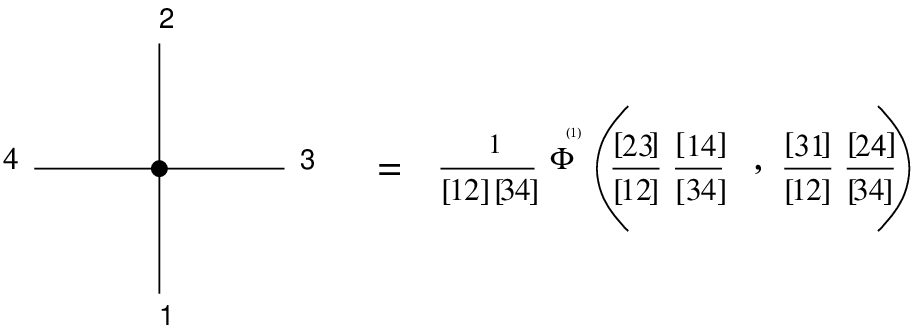,width=\linewidth}
\end{minipage}
\caption{\footnotesize New leg to fourth point  appears due to Jacobian of conformal substitution in the internal vertex}
\label{di7}
\end{figure}

Making the conformal transformation on the l.h.s. and on the r.h.s. of equation in Fig.(\ref{di6}),  
\begin{eqnarray*}
\int~d^4x~ \frac{1}{[1x][2x][3x]} = \frac{1}{[12]} \Phi^{(1)}\le \frac{[23]}{[12]},\frac{[31]}{[12]}\ri.
\end{eqnarray*}
we have 
\begin{eqnarray}
[1'][2'][3']\int~d^4x'\frac{[x']^3}{[1'x'][2'x'][3'x'][x']^4} = \frac{[1'][2']}{[1'2']} \Phi^{(1)}\le \frac{[2'3'][1']}{[1'2'][3']},\frac{[3'1'][2']}{[1'2'][3']}\ri, \label{rev}
\end{eqnarray}
and after shifting the variables 
\begin{eqnarray*}
x' = x''-{x_4}'' \\
{x_1}' = {x_1}'' - {x_4}'', ~~{x_2}' = {x_2}'' - {x_4}'', ~~{x_3}' = {x_3}'' - {x_4}'', 
\end{eqnarray*}
Eq.(\ref{rev}) takes the form 
\begin{eqnarray}
\int~d^4x''\frac{1}{[1''x''][2''x''][3''x''][4''x'']} = 
\frac{1}{[1''2''][3''4'']} \Phi^{(1)}\le \frac{[2''3''][1''4'']}{[1''2''][3''4'']},\frac{[3''1''][2''4'']}{[1''2''][3''4'']}\ri. \label{rev2}
\end{eqnarray}
Omitting double prime symbols we can depict Eq.(\ref{rev2}) in Fig.(\ref{di7}).

The creation of a new fourth factor in the denominator of integral with help of Jacobian of the conformal substitution is in some sense an opposite trick 
to the uniqueness method of Refs. \cite{Unique,Vasiliev:1981dg,Kazakov:1984bw,Vasil}. The uniqueness method is based on elimination of one of 
the three factors in the denominator of the unique integral due to the Jacobian of the conformal substitution. That trick of elimination should be done in 
the Euclidean space in order to avoid possible difficulties with the imaginary part of the causal massless propagators. 
The Euclidean space metrics can be obtained by Wick rotation \cite{Usyukina:1992jd}  from Minkowski metrics.
After the integration according to the uniqueness method the Wick rotation can be done back in order to recover the signature of the Minkowski metrics. 
This means that a small imaginary part should be added to the square of spacetime distances in the denominator. 

The same philosophy can be applied to the constructions done in the present paper. All the formulas and figures should be understood in the Euclidean space. 
All the transformations with creating and eliminating of a new propagator are done in the Euclidean space in analogy with the  uniqueness method. 
The Minkowski metrics signature can be recovered with help of the Wick rotation in each of the integrations in the internal vertices of graph. 
A small imaginary part should be added to the square of spacetime distances which are parts of the arguments of the UD functions 
(the arguments are fractions of the spacetime distances) in the formulas of the present paper on the r.h.s. of the Figures. 

For example, Fig.(\ref{di6}) and Fig.(\ref{di7}) are related in the Euclidean space before the Wick rotation due to the conformal substitution. 
After the Wick rotation a small imaginary parts appear in the propagators on the l.h.s. and in the numerators and denominators of the fractions of spacetime distances  
on the r.h.s. in the arguments of the UD functions. Both sides of Fig.(\ref{di7}) can be represented as a sum of real and imaginary parts.  
This imaginary part on the r.h.s. of Fig.(\ref{di7})  has been evaluated in Ref.\cite{Duplancic:2002dh}  via Mellin-Barnes transformation. 

\begin{figure}[htb]
\begin{minipage}[h]{1.0\linewidth}
\centering\epsfig{file=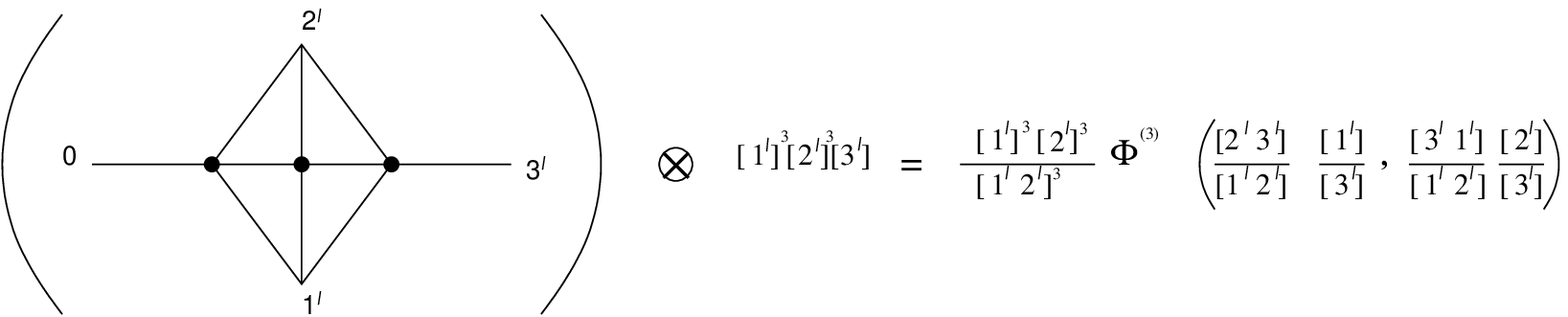,width=\linewidth}
\end{minipage}
\caption{\footnotesize New leg in three-point function due to Jacobian of conformal substitution in the internal vertices}
\label{di8}
\end{figure}

\begin{figure}[ht]
\begin{minipage}[h]{0.8\linewidth}
\centering\epsfig{file=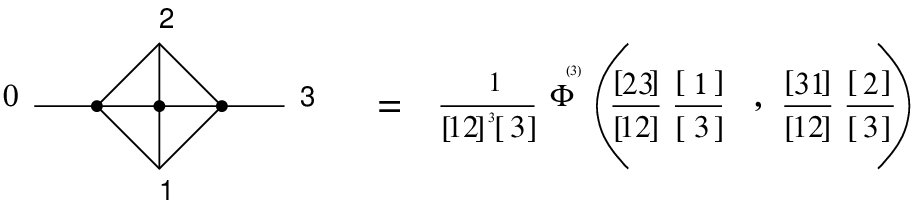,width=\linewidth}
\end{minipage}
\caption{\footnotesize Result of the conformal substitution in the position space}
\label{di9}
\end{figure}

\begin{figure}[ht]
\begin{minipage}[h]{1.0\linewidth}
\centering\epsfig{file=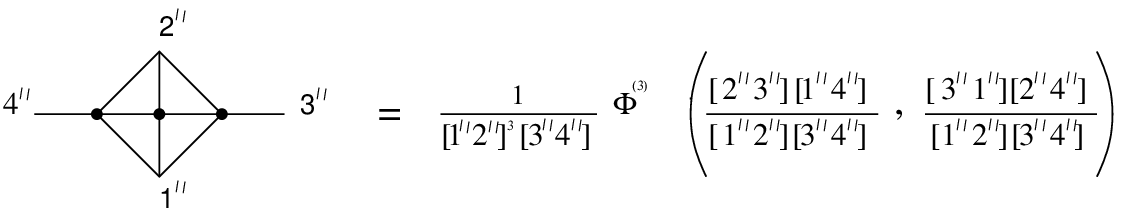,width=\linewidth}
\end{minipage}
\caption{\footnotesize Shift of variables by making use of translation invariance}
\label{di10}
\end{figure}

\begin{figure}[ht]
\begin{minipage}[h]{0.8\linewidth}
\centering\epsfig{file=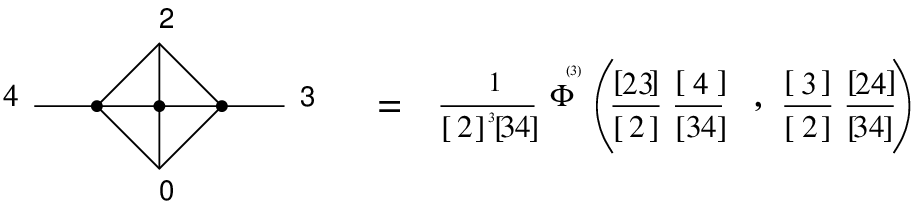,width=\linewidth}
\end{minipage}
\caption{\footnotesize Particular case of Fig.(10)}
\label{di11}
\end{figure}

\begin{figure}[ht]
\begin{minipage}[h]{0.7\linewidth}
\centering\epsfig{file=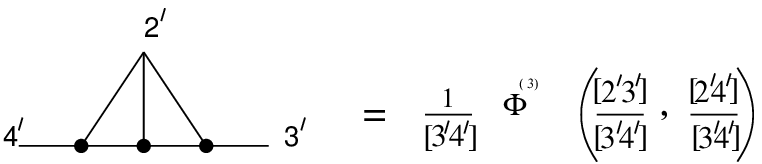,width=\linewidth}
\end{minipage}
\caption{\footnotesize Removing point ``0'' due to Jacobian of conformal substitution in the internal vertices}
\label{di12}
\end{figure}

\begin{figure}[ht]
\begin{minipage}[h]{0.7\linewidth}
\centering\epsfig{file=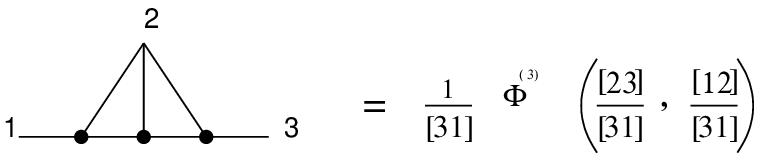,width=\linewidth}
\end{minipage}
\caption{\footnotesize This picture is  Fig.(12) with redefined variables}
\label{di13}
\end{figure}

\begin{figure}[ht]
\begin{minipage}[h]{0.8\linewidth}
\centering\epsfig{file=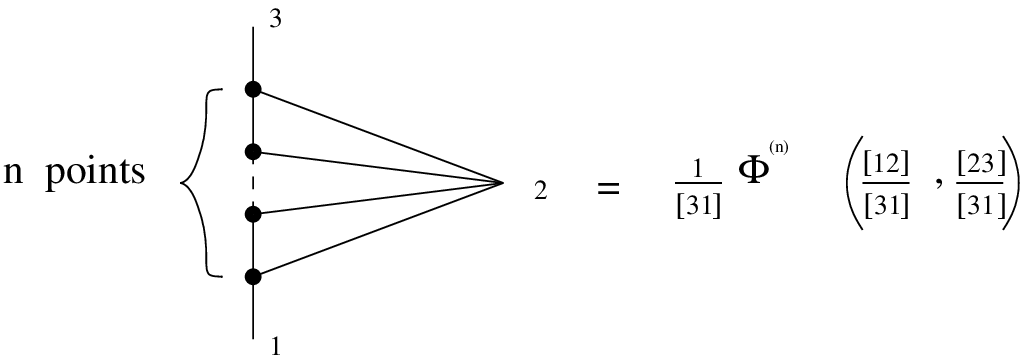,width=\linewidth}
\end{minipage}
\caption{\footnotesize The result for Fig.(13) generalized for arbitrary number of internal vertices}
\label{di14}
\end{figure}

\section{Conformal substitution for an arbitrary number $n$}

As an example, we consider transformation of the diagram depicted in Fig.(\ref{di5}) with three points of integration. 
We generalize the trick of Eq.(\ref{rev}) and  Eq.(\ref{rev2}) and obtain the formula of Fig.(\ref{di8}).
After making simple algebra and removing the prime symbols the formula of Fig.(\ref{di8}) transforms to formula of  Fig.(\ref{di9}).

In order to obtain Fig.(\ref{di10}) we make another substitution (shift of integration variables)
\begin{eqnarray*}
{y_i} = {y_i}''-{x_4}'' \\
{x_1} = {x_1}'' - {x_4}'',~~{x_2} = {x_2}'' - {x_4}'',~~{x_3} = {x_3}'' - {x_4}'',
\end{eqnarray*}
in which ${y_i}$ (or ${y_i}''$) are variables of integration in the internal vertices of Fig.(\ref{di9})  (or Fig.(\ref{di10})), $i=1,2,3.$ 
Removing the prime symbols and putting $x_1=0$ (as a particular case for which equation of Fig.(\ref{di10}) is valid), we obtain Fig.(\ref{di11}).
By making conformal substitution, we remove point ``0'' and propagators connecting  it to other internal vertices due to Jacobians of the 
conformal substitutions in each internal vertex of integration of  Fig.(\ref{di11}).   This results in  Fig.(\ref{di12}). 
Removing the prime symbols and replacing $x_4$ with $x_1$ we obtain Fig.(\ref{di13}) from   Fig.(\ref{di12}).
The same steps from Fig.(\ref{di8}) to Fig.(\ref{di13}) can be reproduced for an arbitrary number $n$ of points corresponding to internal vertices of integration 
in the line connecting points $x_1$ and $x_3$ ($n$ loops in momentum space), and we obtain as a conclusion the relation depicted in Fig.(\ref{di14}).

The diamond diagram appears at intermediate steps from Fig.(\ref{di8}) till Fig.(\ref{di11}) from the dual image of three-point diagram in the m.s.r.
However, we did not use the result of  Refs. \cite{Usyukina:1992jd,Usyukina:1993ch,Drummond:2006rz,Isaev:2003tk} for the diamond diagram. 
In Fig.(\ref{di9}) the result for diamond diagram is obtained as a consequence of the change of variables in the arguments of UD function which is done 
together  with the integral substitution on the l.h.s. of Fig.(\ref{di8}).

\section{Conclusion}

Equation in Fig.(\ref{di14}) combined with iterative definition (\ref{iter}) of the UD function produces relations between  the triangle ladder diagrams 
and the UD functions which were found in Ref.\cite{Kondrashuk:2008xq}. As it was shown, the result $T_n([12],[23],[31])$ 
for $n$-rungs triangle ladder diagram in the position space satisfies the equations 
\begin{eqnarray}
\le \pd_{(2)}^2 \ri^{n-1}~T_n([12],[23],[31]) = \frac{(-4\pi^2)^{n-1}}{[31]} \Phi^{(n+1)}\le \frac{[12]}{[31]},\frac{[23]}{[31]}\ri, \no\\
\le \pd_{(2)}^2 \ri^{n}~T_n([12],[23],[31]) = \frac{(-4\pi^2)^n}{[12][23]} \Phi^{(n)}\le \frac{[12]}{[31]},\frac{[23]}{[31]}\ri, \no\\
\le \pd_{(2)}^2 \ri^{n}~T_{n+1}([12],[23],[31]) = \frac{(-4\pi^2)^n}{[31]} \Phi^{(n+2)}\le \frac{[12]}{[31]},\frac{[23]}{[31]}\ri. \label{UD}
\end{eqnarray}
These relations show that the Fourier transform of UD function with number $n$ is the same UD function with number $n,$
\begin{eqnarray}
\frac{1}{[31]^2} \Phi^{(n)}\le \frac{[12]}{[31]},\frac{[23]}{[31]}\ri = \frac{1}{(2\pi)^4}\int~d^4p_1d^4p_2d^4p_3 ~ \delta(p_1 + p_2 + p_3) \times\no\\
\times e^{ip_2x_2} e^{ip_1x_1} e^{ip_3x_3} \frac{1}{(p_2^2)^2} \Phi^{(n)}\le \frac{p_1^2}{p_2^2},\frac{p_3^2}{p_2^2}\ri. \label{SF}
\end{eqnarray}
The explicit form of the function is given in Refs.\cite{Davydychev:1992xr,Usyukina:1992jd,Usyukina:1993ch},
\begin{eqnarray}
\Phi^{(n)}\le x,y\ri = -\frac{1}{n!\L}\sum_{j=n}^{2n}\frac{(-1)^j j!\ln^{2n-j}{(y/x)}}{(j-n)!(2n-j)!}\left[\Li_j\le-\frac{1}{\rho x} \ri - \Li_j(-\rho y)\right], \label{explicit}
\end{eqnarray}
in which 
\begin{eqnarray*}
\rho = \frac{2}{1-x-y+\L}, ~~~~ \L = \sqrt{(1-x-y)^2-4xy}.
\end{eqnarray*}
Looking at this explicit form of the UD functions $\Phi^{(n)}\le x,y\ri$ it is difficult to imagine that they do not change their form under the Fourier transformation. 
However, the relations  (\ref{UD}) were derived in Ref. \cite{Kondrashuk:2008xq} in diagrammatic way without any use of 
the explicit expression in terms of polylogarithms (\ref{explicit}).  To our knowledge, this is the unique example of the non-trivial functions (actually infinite set of functions) 
possessing such a property. 

Nevertheless, Eq. (\ref{UD})  was an occasional finding of a family of functions with such a property of invariance with respect to Fourier transform. Other examples can be 
constructed manually. Indeed,  Eqs. (\ref{SF}) can be cross-checked via MB transformation \cite{Allendes:2009bd}. In the demonstration of Ref.\cite{Allendes:2009bd} 
the explicit Mellin-Barnes image of the UD functions does not play any role and in principle can be replaced with any function of two complex variables 
with sufficiently good behaviour at complex  infinity, which possesses nontrivial set of left and right residues. For example, Eq.(\ref{SF}) can be proved  via MB transformation  
for the first UD function $\Phi^{(1)}\le x,y\ri$ which does not appear in the chain of Ref. \cite{Kondrashuk:2008xq}.  That chain of transformation 
starts with the second UD function $\Phi^{(2)}\le x,y\ri.$  Another example would be a combination of Appell's hypergeometric functions 
of Ref.\cite{Davydychev:1992xr} which corresponds to a three-point integral of three scalar propagators with arbitrary powers 
in denominator.

\subsection*{Acknowledgments}

This work is supported by Fondecyt (Chile) grants  \#1040368, \#1050512 and by DIUBB grant (UBB, Chile) \#082609.  
This paper is based on I.K.'s talks at   ``High energy physics in the LHC era'' conference, Valparaiso (Chile), The XVIIth Oporto Meeting on Geometry, Topology and Physics 
``Mathematical aspects of quantum field theory'', Oporto  (Portugal) , and at the seminars at Departamento de Fisica, Universidad de Concepci\'on (Chile),
Theoretical Physics department, Karlsruhe University (Germany), Department of Mathematics, University of Bergen (Norway),
DMFA seminar, UCSC, Concepci\'on and IMAFI seminar, Universidad de Talca (Chile). He is grateful to organizers of all these  events for opportunity to present the results.


\begin{thebibliography}{99}



\bibitem{Davydychev:1992xr}
A.~I.~Davydychev, ``Recursive algorithm of evaluating vertex type Feynman integrals,''
  J.\ Phys.\ A {\bf 25}, 5587 (1992).



\bibitem{Usyukina:1992jd}
  N.~I.~Usyukina and A.~I.~Davydychev, ``An Approach to the evaluation of three and four point ladder diagrams,''
  Phys.\ Lett.\  B {\bf 298} (1993) 363.



\bibitem{Usyukina:1993ch}
  N.~I.~Usyukina and A.~I.~Davydychev,
  ``Exact results for three and four point ladder diagrams with an arbitrary number of rungs,''
  Phys.\ Lett.\  B {\bf 305} (1993) 136.



\bibitem{Kondrashuk:2008ec} 
I.~Kondrashuk and A.~Kotikov, ``Fourier transforms of UD integrals,''  arXiv:0802.3468 [hep-th], 
Birkhauser book series ``Trends in Mathematics'', volume``Analysis and Mathematical Physics'', 
B. Gustafsson and A. Vasil'ev (Eds), (2009) Birkhauser Verlag, Basel, Switzerland, 337-348



\bibitem{Kondrashuk:2008xq}
I.~Kondrashuk and A.~Kotikov, ``Triangle UD integrals in the position space,'' JHEP {\bf 0808} (2008) 106 [arXiv:0803.3420 [hep-th]].





\bibitem{Slavnov:1972fg}
A.~A.~Slavnov, 
``Ward Identities In Gauge Theories,'' 
Theor.\ Math.\ Phys.\  {\bf 10} (1972) 99
[Teor.\ Mat.\ Fiz.\  {\bf 10} (1972) 153].




\bibitem{Taylor:1971ff}
J.~C.~Taylor, 
``Ward Identities And Charge Renormalization Of The Yang-Mills Field,''
Nucl.\ Phys.\ B {\bf 33} (1971) 436.




\bibitem{Slavnov:1974dg}
A.~A.~Slavnov,
``Renormalization Of Supersymmetric Gauge Theories. 2. Nonabelian Case,''
Nucl.\ Phys.\ B {\bf 97} (1975) 155.



\bibitem{Faddeev:1980be}
L.~D.~Faddeev and A.~A.~Slavnov,
``Gauge Fields. Introduction To Quantum Theory,''
Front.\ Phys.\  {\bf 50}, 1 (1980)
[Front.\ Phys.\  {\bf 83}, 1 (1990)];
``Introduction to quantum theory of gauge fields'', Moscow, Nauka, (1988).



\bibitem{Lee:1973hb}
B.~W.~Lee,
``Transformation Properties Of Proper Vertices In Gauge Theories,''
Phys.\ Lett.\ B {\bf 46} (1973) 214.



\bibitem{Zinn-Justin:1974mc}
J.~Zinn-Justin,
``Renormalization Of Gauge Theories,''
SACLAY-D.PH-T-74-88,
{\it Lectures given at Int. Summer Inst. for Theoretical Physics, 
Jul 29 - Aug 9, 1974, Bonn, West Germany}.







\bibitem{Cvetic:2004kx}
G.~Cveti\v{c}, I.~Kondrashuk and I.~Schmidt, ``Effective action of dressed  mean fields for N = 4 super-Yang-Mills theory,''
Mod.\ Phys.\ Lett.\ A {\bf 21} (2006) 1127 
[hep-th/0407251].





\bibitem{Kondrashuk:2004pu}
I.~Kondrashuk and I.~Schmidt, ``Finiteness of N = 4 super-Yang-Mills 
effective action in terms of dressed N = 1 superfields,'' arXiv:hep-th/0411150




\bibitem{Kang:2004cs}
 K.~Kang and I.~Kondrashuk, ``Semiclassical scattering amplitudes of dressed gravitons,'' hep-ph/0408168.





\bibitem{Cvetic:2006kk}
G.~Cveti\v{c}, I.~Kondrashuk and I.~Schmidt,
``On the effective action of dressed mean fields for N = 4 super-Yang-Mills theory,'' 
in {\it Symmetry, Integrability and Geometry: Methods and Applications,} SIGMA (2006) 002
[math-ph/0601002].




\bibitem{Cvetic:2006iu}
G.~Cveti\v{c}, I.~Kondrashuk, A.~Kotikov and I.~Schmidt, ``Towards the two-loop Lcc vertex in Landau gauge,'' 
Int.\ J.\ Mod.\ Phys.\  A {\bf 22} (2007) 1905
[arXiv:hep-th/0604112].



\bibitem{Cvetic:2007fp} G.~Cveti\v{c} and I.~Kondrashuk, ``Further results for the two-loop Lcc vertex in the Landau gauge,''
JHEP {\bf 0802} (2008) 023 
[arXiv:hep-th/0703138].



\bibitem{Cvetic:2007ds} G.~Cveti\v{c} and I.~Kondrashuk, ``Gluon self-interaction in the position space in Landau gauge,''
Int. \ J. \ Mod.  \ Phys.  A {\bf 23} (2008) 4145 
[arXiv:0710.5762 [hep-th]].




\bibitem{Allendes:2009bd}
  P.~Allendes, N.~Guerrero, I.~Kondrashuk and E.~A.~Notte~Cuello, ``New four-dimensional integrals by Mellin-Barnes transform,''
  arXiv:0910.4805 [hep-th].




\bibitem{Cvetic:2002in}
G.~Cveti\v{c}, I.~Kondrashuk and I.~Schmidt, 
``QCD effective action with dressing functions: Consistency checks inperturbative regime,'' 
Phys.\ Rev.\ D {\bf 67} (2003) 065007
[hep-ph/0210185].



\bibitem{Kondrashuk:2000br} 
I.~Kondrashuk, 
``The solution to Slavnov-Taylor identities in D4 N = 1 SYM,'' 
JHEP {\bf 0011}, 034 (2000)
[hep-th/0007136].


\bibitem{Isaev:2003tk}
  A.~P.~Isaev, ``Multi-loop Feynman integrals and conformal quantum mechanics,''
  Nucl.\ Phys.\  B {\bf 662} (2003) 461
  [arXiv:hep-th/0303056].





\bibitem{Drummond:2006rz}
 J.~M.~Drummond, J.~Henn, V.~A.~Smirnov and E.~Sokatchev, ``Magic identities for conformal four-point integrals,'' JHEP {\bf 0701} (2007) 064 [arXiv:hep-th/0607160].



\bibitem{Kazakov:1986mu}
 D.~I.~Kazakov and A.~V.~Kotikov,  ``The method of uniqueness ``Multiloop  calculation  in QCD''  Theor.\ Math.\ Phys.\  {\bf 73} (1988) 1264
[Teor.\ Mat.\ Fiz.\  {\bf 73} (1987) 348];




\bibitem{Kazakov:1984bw}
 D.~I.~Kazakov,  ``Analytical Methods For Multiloop Calculations: Two Lectures On The Method Of Uniqueness,'' JINR-E2-84-410.




\bibitem{Vasil} A.N. Vasiliev, ``The field theoretic renormalization group in critical behaviour theory and stochastic dynamics'',  
St. Petersburg Institute of Nuclear Physics Press, 1998. 



\bibitem{Unique} M.~D'Eramo, L.~Peliti and G. Parisi,  ``Theoretical Predictions for Critical Exponents at the $\L$-Point of Bose
Liquids,'' Lett. Nuovo Cimento 2 (1971) 878.



\bibitem{Vasiliev:1981dg} A.~N.~Vasiliev, Y.~M.~Pismak and Y.~R.~Khonkonen, ``1/N Expansion: Calculation 
Of The Exponents Eta And Nu In The Order 1/N**2 For Arbitrary Number Of Dimensions,'' Theor.\ Math.\ Phys.\  {\bf 47} (1981) 465
[Teor.\ Mat.\ Fiz.\  {\bf 47} (1981) 291].


\bibitem{Duplancic:2002dh} G.~Duplan\v{c}i\'c and B.~Ni\v{z}i\'c, ``IR finite one-loop box scalar integral with massless internal lines,'' Eur.\ Phys.\ J.\  C {\bf 24} (2002) 385
[arXiv:hep-ph/0201306].
 



\end{thebibliography}
\end{document}